# Scaling Fields in the Two-Dimensional Abelian Sandpile Model

Stéphane Mahieu and Philippe Ruelle[*]
*Université catholique de Louvain*
*Institut de Physique Théorique*
*B–1348    Louvain-la-Neuve, Belgium*
(October 24, 2018)

We consider the isotropic two-dimensional abelian sandpile model from a perspective based on two–dimensional (conformal) field theory. We compute lattice correlation functions for various cluster variables (at and off criticality), from which we infer the field–theoretic description in the scaling limit. We find a perfect agreement with the predictions of a $c = -2$ conformal field theory and its massive perturbation, thereby providing direct evidence for conformal invariance and more generally for a description in terms of a local field theory. The question of the height 2 variable is also addressed, with however no definite conclusion yet.

## I. INTRODUCTION

Sandpile models have been invented by Bak, Tang and Wiesenfeld [1] as prototypical examples for a class of models which show self–organized criticality. The main peculiarity of these models is that they possess a dynamics which drives them to a critical regime, robust against various perturbations. The ineluctable criticality as well as the robustness of these specific dynamics could provide a universal explanation of the ubiquity of power laws in natural phenomena. Various physical situations have been discussed along this idea, see the recent books [2,3].

Sandpile models are among the simplest models showing self–organized criticality. Although their physical relevance can be questioned, it is believed that they have all the features that should be present in more complicated/physical models. Therefore they constitute a useful playground where the most important features can be understood.

One of the most interesting models is the two–dimensional unoriented abelian sandpile model (ASM) [1], which we first briefly recall (recent reviews are [4,5]). The model is defined on a $L \times M$ square lattice. At each site $i$, we assign a random variable $h_i$, taking its values in the set $\{1, 2, 3, 4\}$. We think of $h_i$ as a height variable, which counts the number of grains of sand at $i$. Thus a sand configuration is specified by a set of values $\{h_i\}_i$ of the height variables. A configuration is stable if all $h_i \leq 4$, and unstable if $h_i > 4$ for one or more sites. The number of stable configurations is equal to $4^{LM}$.

The discrete dynamics of the model takes a stable configuration $\mathcal{C}_t$ at time $t$ to another stable configuration $\mathcal{C}_{t+1}$, and is defined in two steps. First step is the addition of sand: one grain of sand is dropped on a randomly chosen site of $\mathcal{C}_t$, and this produces a new configuration $\mathcal{C}'_t$. The second step is the relaxation to $\mathcal{C}_{t+1}$. If $\mathcal{C}'_t$ is stable, we simply set $\mathcal{C}_{t+1} = \mathcal{C}'_t$. If not, the site where $h_i > 4$ topples: it looses 4 grains of sand, and each of its neighbours receives one grain, something we write in the form $h_j \to h_j - \Delta_{ij}$ for all sites $j$, with $\Delta$ the discrete laplacian. In the process, one neighbour can have its height $h > 4$, in which case it too topples: it looses 4 grains of sand, each of its neighbours receiving one grain. And so on for each site which has a height $h > 4$, until we reach a stable configuration. $\mathcal{C}_{t+1}$ is then set equal to this new stable configuration. The relaxation process is well–defined: it always stops (sand can leave the system at the boundaries) and produces the same result $\mathcal{C}_{t+1}$ independently of the order in which the topplings are performed (the abelian property).

One can let an initial distribution over the stable configurations evolve in time according to the dynamics, and examine its time limit. Under mild assumptions, one shows [6] that all initial distributions converge to a well-defined and unique distribution $P^*$, call the SOC (for self–organized critical) state. The theory of Markov chains and the abelian property allow for a complete characterization of it: $P^*$ is uniform on the set $\mathcal{R}$ of so–called recurrent configurations, and is zero elsewhere (the transient configurations). The number of recurrent configurations is $|\mathcal{R}| = \det \Delta \sim (3.21)^{LM}$, with $\Delta$ the discrete laplacian on the $L \times M$ lattice with open boundary conditions. Although the counting of recurrent configurations is easy, the criterion which actually decides whether a given stable configuration is recurrent or transient is well–known [6,7] but hard and non–local: in a generic case, one has to scan the whole configuration in order to decide whether it is recurrent or not. Explicit calculations are therefore difficult (and few).

From the point of view of critical systems and conformal field theory, one is interested in the thermodynamic limit $\lim_{L,M\to\infty} P^*$. The result should be a probability measure on the space of spatially unbounded configurations, or

---

equivalently on the infinite collection of random variables $h_i$. Despite the fact that these variables are strongly coupled —the couplings are even non–local because of the recurrence condition—, their correlation functions seem to be of the usual, local form. In the scaling limit, one could therefore hope to recover a local field theory.

There are indications that indeed a conformal field theory emerges, like in ordinary critical, equilibrium lattice models. In [7], a connection with spanning trees was established, which suggests a relationship with the $q = 0$ limit of the $q$–state Potts model, hence with a $c = -2$ conformal field theory, a value confirmed by the calculation of the universal finite size correction to the free energy on a finite strip [7]. The 2–site probability $\text{Prob}[h_i = h_j = 1]$ was shown in [8] to decay algebraically, with an exponent that can be easily accomodated in a $c = -2$ free grassmanian scalar field theory [9]. Also the 2–site probabilities for height variables on the boundary of a half–plane domain have been computed in [10,11], and show the same algebraic fall–off as the height 1 variables in the bulk.

Beyond these concordant elements, no systematic investigation in the sandpile model has been made, to our knowledge, which can solidly confirm the connection with a $c = -2$ conformal field theory. It is our purpose to provide a more explicit link between the two. We do this by computing multi–site probabilities of various height variables, and by comparing them with the conformal predictions. More specifically, we compute the scaling limit of the 2–, 3– and 4–site correlations of height 1 variables, but also of other lattice variables, namely finite subconfigurations that can be handled by the technique developed in [8].

In fact, we compute these correlations in an off–critical extension of the abelian sandpile model. We evaluate them in the scaling regime, extract the scaling limit, and then establish a correspondence with a field theory. In this way we strengthen the field–theoretic connection away from criticality, by relating a massive perturbation of the ASM to the massive extension of the $c = -2$ fermionic field theory. One can therefore probe more deeply the structure of both pictures, leaving little doubt about the identifications that are to be made.

The conclusion these calculations allow us to draw is that the $c = -2$ theory, and its massive extension, seems to provide a field–theoretic description of the height profile of the sandpile model. At least for the cluster variables examined in this paper, this is a statement that we could verify explicitly. Other important spatial, non–dynamical features of the SOC state must be studied. These include boundary features and avalanche distributions. The latters are undoubtedly much more difficult to handle, because they lie at a higher level of non–locality than the height variables, since they depend on height values in unbounded regions. Whether they can be accounted for by the non–local sectors of the $c = -2$ conformal theory remains a largely open question.

## II. LATTICE CALCULATIONS IN THE SANDPILE MODEL

As recalled above, explicit calculations in the bulk of the lattice are notoriously hard, because of the non–local nature of the SOC state (probability measure) $P^*$.

All four 1–site probabilities $\text{Prob}[h_i = a]$, for $a = 1, 2, 3, 4$, have been computed exactly in the thermodynamic limit, but already the calculation for $a \geq 2$ [12] is formidably more complicated than for $a = 1$ [8]. The only 2–site probability that has been computed is again for the unit height variables [8].

The technique used to compute the correlation of two unit height variables is a particular case of a beautiful idea put forward by Majumdar and Dhar [8]. It is based on the important notion of forbidden subconfigurations (FSC), and its relation to recurrent configurations. A cluster $F$ of sites, with its heights $h_i$, is a FSC if, for each site $j \in F$, the number of sites in $F$ and connected to $j$ is bigger or equal to $h_j$. Simple examples of FSCs are two adjacent 1's (11), a linear arrangement (121), or a cross–shape arrangement with four 1's surrounding a central site with any height value. A configuration is then recurrent if and only if it contains no FSC [6,7].

The idea used in [8] allows to compute the probability of occurrence, in the SOC state, of any cluster that becomes an FSC if any of its heights is decreased by one unit. A simple case is a height 2 next to a height 1, but more examples are given below in Figure 1. Following [13], let us call them weakly allowed cluster variables.

Let $S$ be such a cluster. The authors of [8] show how one can define a new sandpile model, with its own toppling rules (and a new matrix $\Delta'$), such that the number of its recurrent configurations is the number, in the original model, of recurrent configurations which contain $S$. From this, a simple determinantal formula follows, $\text{Prob}(S) = \det \Delta' / \det \Delta$. Because the new sandpile model is obtained by modifying the original one in the region localized around $S$, the ratio of the two determinants reduces to a finite determinant, even for an infinite lattice.

This technique has been used to compute the probabilities of various subconfigurations, like those in Figure 1 below. The simplest one is the cluster reduced to one site, with height equal to 1. In this case, the new model is obtained by changing the toppling rules at 4 sites (the height 1 and three neighbours). A 4–by–4 determinant then yields $P(1) \equiv \text{Prob}[h_i = 1] = \frac{2}{\pi^2}(1 - \frac{2}{\pi}) \sim 0.074$. Allowing for disconnected clusters leads to multi–site correlations such as the 2–site correlation of unit heights:

$$\text{Prob}[h_i = h_j = 1] = P(1)^2 \left[1 - \frac{1}{2r^4} + \ldots\right], \qquad r = |i - j| \gg 1. \tag{2.1}$$

It was also remarked in [8] that more general clusters —for instance, a single site with height equal to 2— can be handled using the same ideas, but the corresponding probabilities become infinite series, the terms of which involve weakly allowed clusters, of increasing size. Unfortunately, these series seem to be slowly convergent.

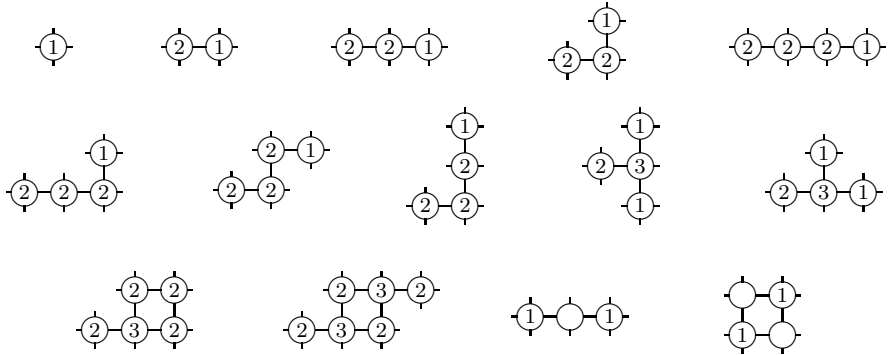

FIG. 1. On the first two lines are shown the ten smallest weakly allowed cluster variables, up to orientations, which contain no more than four sites. Taking the different orientations into account makes a total of 57 clusters of weight smaller or equal to 4. In addition, calculations involving the four clusters on the last line will be considered in the text. All these clusters will be numbered $S_0$ to $S_{13}$ from left to right and top to bottom. The reason for including the last two clusters is explained in Section VII.

In general, the way the original sandpile model is modified is by removing some of the bonds linking $S$ to its nearest neighbourhood, and at the same time, by reducing the threshold at which the sites become unstable (4 in the original model), so that the threshold at every site remains equal to its connectivity. These modifications affect all the sites of $S$, plus a certain number of sites which are nearest neighbours of $S$. All together they form a set we call $M_S$, the cardinal of which depends on the shape of $S$. The new toppling matrix is then given by $\Delta' = \Delta + B$, where the symmetric matrix $B$ has entries $B_{ij} = 1$ if the bond linking $i$ to $j$ has been removed, $B_{ii} = -n$ if $n$ bonds off the site $i$ have been removed, and is zero otherwise. Then the probability of $S$ (in the original model) is

$$P(S) = \frac{\det \Delta'}{\det \Delta} = \det (\mathbb{I} + GB) = \det (\mathbb{I} + GB)\big|_{M_S}. \tag{2.2}$$

Because $B$ is zero outside the finite set $M_S$, the determinant is finite, in fact of size $|M_S|$, but requires the knowledge of the Green function $G \equiv \Delta^{-1}$ of the laplacian, at all sites belonging to $M_S$.

In the above example where $S$ is just one site with a height equal to 1, the modifications can be pictorially described as follows:

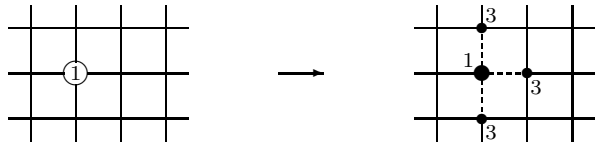

The dashed segments represent the removed bonds, and the numbers on the right lattice indicate the thresholds at which the sites become unstable and topple.

In fact, in the modified lattice shown on the right, the only site to which the 1 is connected has a height bigger or equal to 2. So one could as well decrease its height and its threshold by 1, and remove the connection. In this way, the site with a height originally equal to 1 is completely cut off from the rest of the lattice, defining a different modification of the original ASM[1]. Either of them can be used to compute correlations involving heights 1.

Correspondingly the matrix $B$ that specifies the modifications is a 4–by–4 or a 5–by–5 matrix given by (in an obvious ordering)

$$B = \begin{pmatrix} -3 & 1 & 1 & 1 \\ 1 & -1 & 0 & 0 \\ 1 & 0 & -1 & 0 \\ 1 & 0 & 0 & -1 \end{pmatrix} \quad \text{or} \quad \begin{pmatrix} -3 & 1 & 1 & 1 & 1 \\ 1 & -1 & 0 & 0 & 0 \\ 1 & 0 & -1 & 0 & 0 \\ 1 & 0 & 0 & -1 & 0 \\ 1 & 0 & 0 & 0 & -1 \end{pmatrix}. \tag{2.3}$$

For bigger clusters, there is a fair amount of ambiguity in the way the modifications are made in order to freeze the cluster heights to what we want. These modifications can affect regions of different sizes, and so can be more or less computationally convenient. The least economical solution is the analog of the second modification explained above for the unit height. It is also the easiest to describe: one simply cuts the cluster off the rest of the lattice, removing all bonds inside the cluster and all connections between the cluster and the outside lattice. There are many other choices of intermediate efficiency. For the second cluster in Figure 1 for instance, namely a 2 next to a 1, one may consider the following three modifications (among others):

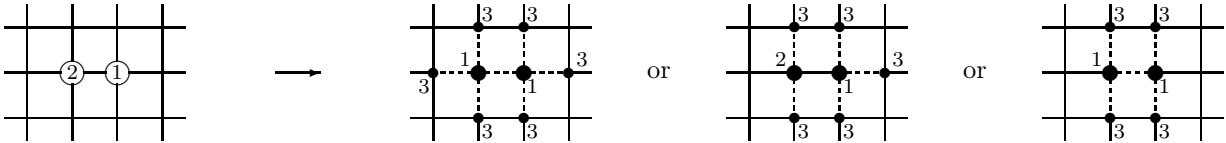

with corresponding $B$ matrices of dimension 8, 7 or 6. For bigger clusters, the difference can be computationally noticeable, and so choosing the modifications which affect the smallest possible region makes the calculation of determinants easier[2]. So one should cut as few links as possible, a prescription which makes sure that the modified ASM remains conservative where the original one is: the removal of a bond off one site is always accompanied by the lowering by 1 of the threshold at that site, or equivalently the $B$ matrix has row and column sums equal to 0.

When the cluster $S = \cup_k S_k$ is disconnected, the matrix $B$ is the direct sum of submatrices $B_k$. The probability Prob($S$) (the correlation of the subparts $S_k$) involves the Green function $G(i,j) = G(0, i-j)$ at all sites $i, j$ of $S$, and thus depends on the relative locations and orientations of the various $S_k$'s, and in particular on their separation distances. As the original sandpile model is invariant under lattice translation, the probabilities retain the translation invariance. For $S$ containing two heights equal to 1, separated by a distance $r$, the evaluation of the 8–by–8 determinant yields the dominant term $r^{-4}$ given in (2.1), independently of the angular distance of the two sites.

Precisely in the case $S$ is disconnected and contains different pieces separated by large distances, a simple but important observation can be made. Because the probability of $S$ is going to depend on the Green function $G(z_k, z_{k'}) \sim \log |z_k - z_{k'}|$ evaluated at points where the subparts are located, one could expect at first sight a logarithmic dependence in the separation distances. However due to the property that sand in conserved in the modified ASM, the probability in fact only depends on the derivatives (or finite differences) of the Green function. This evacuates the logarithmic dependences and turn all correlations into rational series in the various distances $(z_k - z_{k'})$.

## III. THE MASSIVE SANDPILE MODEL

The previous section summarized the calculation of correlations of cluster variables in the standard ASM. Even though it is critical, and self–organized in the dynamical sense, one can drive it off criticality by switching on relevant perturbations. There are various ways of doing it, but one of the simplest is to add dissipation, whose rate is controlled by a parameter $t$. In effect this introduces a mass $m \sim \sqrt{t}$, or equivalently a non–zero reduced temperature $t$. The resulting model can be described as a massive (or thermal) perturbation of the massless, critical sandpile model. For the purpose of comparing the correlations in the ASM with those of a local field theory, the inclusion of some neighbourhood of the critical point is important as it strengthens the connection.

---

[1]Strictly speaking, in this second modified ASM, the removal of the bond connecting the height 1 to its western neighbour should be supplemented by the creation of a bond connecting the height 1 to a sink site, so that sand brought by seeding can be evacuated. The part of the modifications that affect the sink site plays no role whatsoever, so we may ignore it completely. See the second appendix for a detailed argument.

[2]The reader familiar with the technique knows that these determinants can be reduced by appropriate summations of rows (or columns). The gain in size is equal to the size of the cluster one considers, but it has its prize, because it renders the entries of the reduced determinant more complex. This gain is the same no matter how the ASM is modified.

The way a mass can be introduced in the model is most straightforward, and corresponds to a dissipation of sand each time a toppling takes place. We define the perturbed ASM by its toppling matrix (we suppress the explicit dependence on $x$)

$$\Delta_{i,j} = \begin{cases} x & \text{if } i = j, \\ -1 & \text{if } i \text{ and } j \text{ are nearest neighbours}, \\ 0 & \text{otherwise}. \end{cases} \quad (3.1)$$

The external driving rate of the sandpile remains the same (one grain per unit of time), but the threshold beyond which the sites become unstable is increased from 4 to $x$. As a consequence, the height variables now take the values between 1 and $x$. Each time a site topples, $t = x - 4 \geq 0$ grains of sand are dissipated.

In order to assess the robustness of their SOC features, perturbations of sandpile models have been often discussed, with various and sometimes surprising conclusions, see for instance [14] for a review. Among the many contributions on the subject, the reference [15] has been one of the first attempts to see how the non–conservation of sand in the toppling rules can alter the critical properties of the model. In particular, the massive perturbation defined above in (3.1) corresponds to the globally dissipative model studied in that paper, and for which the authors have found that the avalanche distributions decay exponentially. More recently, the same perturbation was reconsidered in [16], in which the exponential decay of the 2–site probability for unit height variables, our Eq. (4.4) below, was proved.

The advantage of the perturbation (3.1) is that it allows the same calculations as the non–perturbed model, in the way that has been recalled in Section II. One can in particular compute the correlation functions by the same formulae, with however two minor changes. The first one is of course that one uses the massive Green function, with a mass fixed by $\sqrt{x-4}$. The second one concerns the $B$ matrices, that define the modified ASM. Because the height variables now take the values from 1 up to $x$, the diagonal entries of $B$ corresponding to sites of the cluster $S$ must be set equal to $1 - x$, in order to lock the heights into their minimal values. (As a consequence, note that sand is conserved at those sites, in the modified model.)

When doing concrete computations, one needs the value of the Green function at points close to the origin (at sites belonging to the same connected subpart), and at points far from the origin (at sites located in different connected parts). For the former, one uses a development around $t = 0$ (in powers of $t$ with $\log t$ terms), whereas for the latter, one performs a double expansion in inverse powers of the distances, and in (half–integral) powers of the perturbing parameter $t$. For arbitrary positions, this development is cumbersome as it depends also on the angular positions. In the calculations to be presented in the following sections, we have therefore restricted ourselves to configurations of clusters which only require the knowledge of the Green function at points close to a principal or a diagonal axis, for which all useful expressions are collected in an appendix.

The field theory enters as a description of the long distance regime of the ASM correlations (perturbed or not). As usual, this requires at the same time an adjustment of the correlation length, or equivalently of the mass. So we are interested in computing the scaling regime of correlations. To reach it, we take simultaneously the long distance limit $R = \frac{r}{a} \to \infty$ and the critical limit[3] $x - 4 = a^2 M^2 \to 0$ so that the product $\sqrt{x-4} R \to Mr$ defines the effective mass $M$ and the macroscopic distances $r$. The scale $a \to 0$ controls the way the limit is taken, and can be thought of as a lattice spacing.

In the actual calculations of correlation functions, large determinants are needed, with entries given by Green function values, themselves expressed as power series. In the scaling regime $x \sim 4$, it is convenient to expand all matrix entries and the correlations as power series of $\sqrt{t}$. The first non–zero term in a correlation should then be directly related to its scaling limit.

We will finish this section by commenting on the way the calculations have been done, before presenting in the next section the results for the unit height variables.

---

[3]We deliberately take the stand to formally continue all the expressions from integer values of $x$ to arbitrary values $x \geq 4$. Thus we do not define a family of well–defined sandpile models, parametrized by a real number $x \geq 4$. For $x$ rational, this can easily be done, however the limit for $x$ going to zero by rational values is not the usual, original model defined for $x = 4$. We suspect that the model one gets in this specific limit is a model in which the height variables are completely decoupled. See [17] for a related discussion.

Suppose that we want to compute the joined probability for having a certain cluster $S$ at the origin say and an other cluster $S'$ at some site $i$. Each cluster comes with its own set $M_S$ or $M_{S'}$ which contains the sites where the ASM has been modified, the modifications themselves being specified by the matrices $B$ and $B'$. According to the discussion of the previous section, this probability is equal to a determinant

$$\text{Prob}\,[S(0),\ S'(i)] = \det\left(\mathbb{I} + \begin{pmatrix} G_{00} & G_{0i} \\ G_{i0} & G_{ii} \end{pmatrix}\begin{pmatrix} B & 0 \\ 0 & B' \end{pmatrix}\right) = \det\begin{pmatrix} \mathbb{I} + G_{00}\,B & G_{0i}\,B' \\ G_{i0}\,B & \mathbb{I} + G_{ii}\,B' \end{pmatrix}. \tag{3.2}$$

The $G$–blocks collectively denotes Green function values evaluated at two sites belonging to the set $M_S \cup M_{S'}$, with in addition $G_{i0} = (G_{0i})^t$.

We do not want to know the exact value of this determinant, but rather the terms that are dominant in the scaling region, when $i$ is far from the origin.

Using the standard development of a rank $n$ determinant in terms of the matrix entries,

$$\det A = \sum_{\sigma \in S_n} \epsilon(\sigma)\, A_{1,\sigma(1)}\, A_{2,\sigma(2)} \ldots A_{n,\sigma(n)}, \tag{3.3}$$

one may distinguish in (3.2) several types of terms.

The permutations $\sigma$ which do not mix the sites of the cluster $S$ with the sites of $S'$, produce terms which do not depend on the distance $|i|$ separating $S$ from $S'$, and thus contribute a term equal to $[\text{Prob}(S)][\text{Prob}(S')]$.

The other permutations necessarily involve an even number of entries from the off–diagonal blocks. As all such entries are combinations of Green functions, they decay exponentially with the distance. Therefore the 2–point function will be dominated by those terms in the determinant which are quadratic in the off–diagonal Green functions. With the help of the formulae in the Appendix A, these Green functions are all reducible to the single $G(i) = G_{0,i}$, and its derivatives.

The quadratic terms come from the permutations that send one site of the first cluster onto one site of the second cluster, and vice–versa (with possibly two other sites). The contributions of all those permutations can be summed up to yield a formula written in terms of the minors of the diagonal blocks:

$$\text{Prob}\,[S(0),\ S'(i)] = \text{Prob}(S)\cdot\text{Prob}(S') - \text{Tr}\left\{[\text{Mi}(\mathbb{I}+G_{00}B)]^t \cdot (G_{0i}B') \cdot [\text{Mi}(\mathbb{I}+G_{ii}B')]^t \cdot (G_{i0}B)\right\} + \ldots \tag{3.4}$$

Here $\text{Mi}(A) = (-1)^{i+j}\det(A_{\hat{i},\hat{j}})$ denote, up to signs, the minors of $A$ of maximal order ($A_{\hat{i},\hat{j}}$ is the matrix $A$ with the $i$–th row and the $j$–th column removed). Formula (3.4) is exact modulo quartic, sextic, ... terms in the off–diagonal Green functions. It has been used to compute all 2–cluster correlations considered in this article.

In order to determine the dominant term in the perturbing parameter $t$, one still makes an expansion in powers of $\sqrt{t}$ (actually the expansions of elements of the diagonal blocks $G_{00}$ and $G_{ii}$ involve the two kinds of terms $t^{k/2}$ and $t^{k/2}\log t$). To this end, one develops all Green functions around $t=0$ using the formulae of the appendix, and keeps the first non–zero term in the trace. Since the mass $m$ or inverse correlation length is related to $\sqrt{t}$, a first non–zero contribution of the form $t^{(x_1+x_2)/2}\,F\left(G(i\sqrt{t}), G'(i\sqrt{t}), \ldots\right)$ determines the scaling limit of the correlation, hence the corresponding field–theoretic 2–point function, in terms of two fields of scale dimensions $x_1$ and $x_2$, in the usual way. In this respect the presence of a logarithmic singularity $\log t$ in the final result would be the signal that the scaling limit is ill–defined. It turns out, in all the calculations we have performed, that the first non–zero term scales like $t^2$ (yielding $x_1+x_2=4$). Because the off–diagonal terms start off like $\sqrt{t}$ —they are differences of Green functions at neighbouring sites—, it is enough to expand all Green functions up to order $t^{3/2}$, as has been done in the appendix[4] (the 3–cluster correlations require the expansions to order $t^2$).

In fact this procedure has anticipated the results on one point. For the purpose of taking the scaling limit, it is the dominant term in $t$ that we want to determine, while the above procedure determines the dominant term in $t$ among the contributions that are quadratic in the Green functions. So one should also check that no higher than quadratic term in the Green functions brings a $t^2$ contribution. This can easily be done in the following way. Since the off–diagonal terms start off like $\sqrt{t}$, checking the quartic terms is enough, and one can stop the expansion of the off–diagonal blocks to the $\sqrt{t}$ order. To that order, the two blocks $G_{0i}B'$ and $G_{i0}B$ have all their rows identical. Indeed, inside a given column, all entries are finite differences of Green functions evaluated at neighbouring sites, and so differ by second order finite differences of Green functions, i.e. by terms of order $t$. Thus the determinant with $G_{0i} B'$ and $G_{i0} B$ as off–diagonal blocks can be reduced to a determinant where the two off–diagonal blocks have but their first row non zero, and equal to linear combinations of Green functions. Such a determinant has no term that is quartic in the off–diagonal block entries.

---

[4]It would not be the case if the least economical modifications were chosen (the one that cuts the cluster off the rest of the lattice). The $B$ matrix would not have all row sums equal to zero, and consequently the off–diagonal Green functions would have non–zero terms of order 0 in $t$. This would force us to expand everything to order 2 (instead of 3/2) in $t$. So these modifications appear to be doubly inefficient.

To end this section, we give the expansion analogous to (3.4) that pertains to the calculation of 3–cluster correlations. Its proof relies on the same arguments as above regarding permutations. For three clusters rooted at sites $i, j, k$, it reads

$$\begin{aligned}
\text{Prob}\left[S(i), S'(j), S''(k)\right] &= -2\,\text{Prob}(S) \cdot \text{Prob}(S') \cdot \text{Prob}(S'') \\
&+ \text{Prob}(S) \cdot \text{Prob}[S'(j), S''(k)] + \text{Prob}(S') \cdot \text{Prob}[S(i), S''(k)] + \text{Prob}(S'') \cdot \text{Prob}[S(i), S'(j)] \\
&+ \text{Tr}\left\{[\text{Mi}(\mathbb{I} + G_{ii}B)]^t \cdot (G_{ij}B') \cdot [\text{Mi}(\mathbb{I} + G_{jj}B')]^t \cdot (G_{jk}B'') \cdot [\text{Mi}(\mathbb{I} + G_{kk}B'')]^t \cdot (G_{ki}B)\right\} \\
&+ \text{Tr}\left\{[\text{Mi}(\mathbb{I} + G_{ii}B)]^t \cdot (G_{ik}B'') \cdot [\text{Mi}(\mathbb{I} + G_{kk}B'')]^t \cdot (G_{kj}B') \cdot [\text{Mi}(\mathbb{I} + G_{jj}B')]^t \cdot (G_{ji}B)\right\} + \ldots \quad (3.5)
\end{aligned}$$

This formula gives all terms of the determinant that are cubic in the off–diagonal Green functions. They are to be expanded around $t = 0$ as discussed above.

## IV. UNIT HEIGHT VARIABLES

The simplest cluster variable is $S_0$, namely the unit height variable. We give in this section its multi–site correlation functions, in various configurations, as computed along the lines exposed above.

The 1–point function, namely the probability a fixed site has height equal to 1, poses no problem (and is in any case of little interest for the comparison with a field theory). Making everything very explicit for once, it is given by

$$\text{Prob}(S_0) \equiv P(1) = \det\left(\mathbb{I} + \begin{pmatrix} G(0,0) & G(1,0) & G(1,0) & G(1,0) \\ G(1,0) & G(0,0) & G(1,1) & G(2,0) \\ G(1,0) & G(1,1) & G(0,0) & G(1,1) \\ G(1,0) & G(2,0) & G(1,1) & G(0,0) \end{pmatrix} \begin{pmatrix} 1-x & 1 & 1 & 1 \\ 1 & -1 & 0 & 0 \\ 1 & 0 & -1 & 0 \\ 1 & 0 & 0 & -1 \end{pmatrix}\right). \quad (4.1)$$

Here $G(m, n)$ is $(\Delta^{-1})_{i,0}$ for the site $i = (m, n)$, and we have used the symmetries of the Green function. The site ordering is O, N, E and S.

This can be easily computed in terms of complete elliptic functions (see the appendix), although the result is not particularly transparent

$$P(1) = \frac{1}{256}\left[(x-4)G(0,0) - 1\right]\left[x^2 G(0,0) - 16 G(1,1) - (x+4)\right]\left[(x^2-8)G(0,0) - 8G(1,1) - (x-4)\right]^2. \quad (4.2)$$

It goes to $2[2G(1,1) - 2G(0,0) + 1][G(1,1) - G(0,0)]^2 = 2(\pi-2)/\pi^3$ in the limit $x \to 4$.

More interesting is its graph, which shows that $\text{Prob}(S_0)$ increases when $x$ goes away from 4 before falling off algebraically when $x$ keeps growing. The graph of $\text{Prob}(S_0)$ as a function of $x$ is reproduced in Figure 2 as the long–dashed curve.

### A. Two–point correlation

The joined probability for having a 1 at the origin say and an other 1 at a site $i$ is equal to the 8–by–8 determinant

$$\text{Prob}\left[h_0 = h_i = 1\right] = \det\begin{pmatrix} \mathbb{I} + G_{00}B & G_{0i}B \\ G_{i0}B & \mathbb{I} + G_{ii}B \end{pmatrix}, \quad (4.3)$$

where $B$ is the matrix used in (4.1). Because the two clusters are identical, $G_{ii} = G_{00}$.

As mentioned earlier, the expansion of the Green function at arbitrary points tends to be complicated, so we have restricted ourselves to configurations where the Green functions close to a principal axis or a diagonal axis only are required. For the 2–site correlation, this only leaves the two possibilities $i = (m, 0)$ and $i = (m, m)$. Using the formula (3.4), we found in both cases the same answer:

$$\begin{aligned}
&\text{Prob}[h_0 = h_i = 1] - [P(1)]^2 = \\
&\quad - t^2 [P(1)]^2 \left\{\frac{1}{2} K_0''^2(\sqrt{t}|i|) - \frac{1}{2} K_0(\sqrt{t}|i|) K_0''(\sqrt{t}|i|) + \frac{1}{2\pi} K_0'^2(\sqrt{t}|i|) + \frac{1+\pi^2}{4\pi^2} K_0^2(\sqrt{t}|i|)\right\} + \ldots, \quad (4.4)
\end{aligned}$$

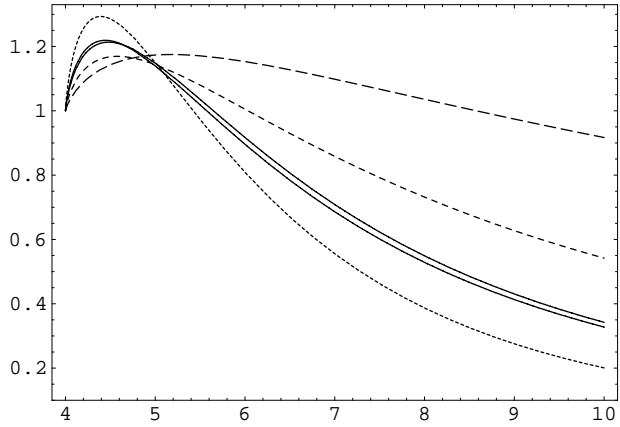

FIG. 2. Probabilities of the clusters $S_0$ up to $S_4$ as functions of the perturbing parameter $x$. In this figure, all probabilities have been normalized to 1 at $x=4$ to prevent some of them to melt into the horizontal axis. The curves from top to bottom refer to $S_0$ (long dashes) down to $S_4$ (shortest dashes). The two solid lines correspond to $S_2$ and $S_3$.

with $|i| = m$ or $\sqrt{2}m$ depending on whether $i$ is real or on the diagonal. The function $K_0$ is the modified Bessel function. Note that the $P(1)$ appearing in the l.h.s. (in the subtraction term) is the off–critical probability, while that in the r.h.s. can be taken to be the critical one.

This formula has a number of instructive and comforting features. The spatial dependence is only through the function $K_0$, that is, the scaling form of the massive lattice Green function. The other functions, denoted $D_i$ or $P_i$ in Appendix A, and representing the lattice corrections to the scaled, continuum Green function, actually do not enter. Moreover the fact that the answer is the same for the two positions of $i$ suggests that the probability is invariant under rotations, in agreement with the rotational invariance of the cluster $S_0$ itself. This is related to the first point, since the functions $D_{i>0}$ and $P_{i>0}$ represent anisotropic terms in the lattice Green function.

Another reassuring feature is that the correlation (4.4) scales like $t^2$, to the dominant order, and that all logarithmic terms $\log t$ have dropped out at that order. This requires massive cancellations because logarithmic terms occur in all entries of the blocks $G_{00}$ and $G_{ii}$, which store Green function values around the origin (see the appendix). We have also checked that (4.4) is exact up to higher order in $t$: all terms of order lower than $t^2$ vanish identically (apart from the zero–th order term $P(1)^2$), and there are no term quartic or higher in the Green function that would contribute a $t^2$ term. Thus (4.4) is exact to order $t^2$.

That the correlation scales like the fourth power of the mass was expected since the critical correlation decays like $|i|^{-4}$ [8]. It is easily recovered from (4.4) by taking the limit $t \to 0$, in which the term in $K_0''^2(|i|\sqrt{t}) \sim \frac{1}{t^2|i|^4}$ is the only one to survive, reproducing the result (2.1).

What the above suggests is that the scaled unit height variable goes over, in the scaling limit, to a massive field $\phi_0$ with scale dimension 2,

$$\lim_{a\to 0} \frac{1}{a^2}\left[\delta\left(h_{z/a} - 1\right) - P(1)\right] = \phi_0(z), \qquad i = \frac{z}{a} \to \infty, \quad t = a^2 M^2 \to 0 \quad \text{with } i\sqrt{t} = Mz, \qquad (4.5)$$

and whose 2–point function reads

$$\langle \phi_0(0)\phi_0(z)\rangle = -M^4\,[P(1)]^2 \left\{\frac{1}{2}K_0''^2(M|z|) - \frac{1}{2}K_0(M|z|)K_0''(M|z|) + \frac{1}{2\pi}K_0'^2(M|z|) + \frac{1+\pi^2}{4\pi^2}K_0^2(M|z|)\right\}. \quad (4.6)$$

### B. Three–point correlation

We made the same calculations for the 3–site probability, using the formula (3.5). The use of the Green functions on the horizontal or the diagonal axis leave essentially two possibilities: either the three insertion points $i$, $j$, and $k$ are aligned, or else they form a isoceles right triangle. In both cases, the probabilities scale like $t^3$, with all logarithms of $t$ cancelled out. The explicit results however differ in these two cases.

When they form a linear arrangement, be it on the horizontal or diagonal axis, the result for the connected probability (i.e. products of lower correlations are subtracted) reads

$$\text{Prob}[h_i = h_j = h_k = 1]_{\text{aligned, connected}} =$$

$$\frac{M^6}{4} [P(1)]^3 \times \Big\{ \big[K_0(12) - K_0''(12)\big] \big[K_0(13) - K_0''(13)\big] \big[K_0(23) - K_0''(23)\big] + K_0''(12) K_0''(13) K_0''(23)$$

$$+ \frac{1}{\pi} \Big[ K_0''(12) K_0'(13) K_0'(23) - K_0'(12) K_0''(13) K_0'(23) + K_0'(12) K_0'(13) K_0''(23) \Big]$$

$$- \frac{1}{\pi^2} \Big[ K_0(12) K_0'(13) K_0'(23) - K_0'(12) K_0(13) K_0'(23) + K_0'(12) K_0'(13) K_0(23) \Big]$$

$$- \frac{1}{\pi^3} K_0(12) K_0(13) K_0(23) \Big\}. \tag{4.7}$$

We have written the answer in the scaled form, that is, after the scaling limit in which the sites $i, j, k$ go over to the macroscopic positions $z_1, z_2$ and $z_3$. The notation $K_0(ij)$ stands for $K_0(M|z_i - z_j|)$.

For the triangular configuration, we chose the insertion points $i = (0,0)$ and $k = (2m, 0)$ to be real, and put $j = (m, m)$ on the diagonal. The result is slightly different in this case, and reads, in the same notations,

$$\text{Prob}[h_i = h_j = h_k = 1]_{\text{triangular, connected}} =$$

$$-\frac{M^6}{4} [P(1)]^3 \times \Big\{ 2 K_0''(12) K_0(13) K_0''(23) - K_0''(12) K_0(13) K_0(23) - K_0(12) K_0(13) K_0''(23)$$

$$+ \frac{1}{\pi} \Big[ \sqrt{2} [K_0''(12) - K_0(12)] K_0'(13) K_0'(23) + \sqrt{2} K_0'(12) K_0'(13) [K_0''(23) - K_0(23)]$$

$$+ K_0'(12) [2 K_0''(13) - K_0(13)] K_0'(23) \Big]$$

$$+ \frac{\sqrt{2}}{\pi^2} \Big[ K_0'(12) K_0'(13) K_0(23) - K_0(12) K_0'(13) K_0'(23) \Big] + \frac{2}{\pi^3} K_0(12) K_0(13) K_0(23) \Big\}. \tag{4.8}$$

Exactly the same result was found, as expected, for the rotated configuration where $i$ is at the origin, $j = (m, 0)$ on the real axis and $k = (m, m)$ on the diagonal.

The same comments as for the 2–site correlation apply here but for one point. If indeed the 3–site probability scaling $\sim t^3$ around the critical point is consistent with the dimension 2 of a unit height variable, one observes that the probabilities themselves vanish in the critical limit ($M \to 0$). Thus the scaling limit of three unit height variables in the usual, unperturbed, ASM vanishes:

$$\lim_{\text{scaling}} \text{Prob}\,[h_i = h_j = h_k = 1]_{x=4, \text{connected}} = 0. \tag{4.9}$$

We have checked this result by using the critical Green functions, and found that the probability for three sites aligned along the real axis

$$\text{Prob}[h_i = h_j = h_k = 1]_{\text{real, connected}} = -\frac{P(1)^2}{\pi^3} \Big[ \frac{1}{z_{12}^3 z_{23}^3 z_{13}^2} - \frac{1}{z_{12}^3 z_{23}^2 z_{13}^3} - \frac{1}{z_{12}^2 z_{23}^3 z_{13}^3} \Big]$$

$$+ \frac{P(1)^3}{8} \Big[ \frac{1}{z_{12}^4 z_{23}^4} + \frac{1}{z_{12}^4 z_{13}^4} + \frac{1}{z_{23}^4 z_{13}^4} \Big] + \frac{3 P(1)^3}{4} \Big[ \frac{1}{z_{12}^4 z_{23}^2 z_{13}^2} + \frac{1}{z_{12}^2 z_{23}^4 z_{13}^2} + \frac{1}{z_{12}^2 z_{23}^2 z_{13}^4} \Big] + \text{higher order} \tag{4.10}$$

indeed decays like a global power $-8$ of the separation distances. Moreover the same calculation for the three sites aligned on the diagonal axis produces different coefficients. Thus the dominant term of the critical lattice 3–point function is not isotropic, contradicting the expected rotational invariance, and so should not survive the scaling limit.

### C. Four–point correlation

Finally, we have also determined the 4–site probability for unit height variables, at the critical point only, as otherwise the number of terms grows quickly. So in this case we have used throughout the calculations the expansions at $x = 4$ of the Green functions, also given in the appendix.

We have examined two different arrangements of the insertion points, when they are all aligned on the real axis, and when they lie at the vertices of a square.

When they are all aligned on the real axis, the connected 4–site probability takes a very simple form, at the dominant order,

$$\text{Prob}[h_i = h_j = h_k = h_l = 1]_{\substack{\text{real} \\ \text{connected}}} = -\frac{P(1)^4}{4} \left\{ \frac{1}{(z_{12}z_{34}z_{13}z_{24})^2} + \frac{1}{(z_{13}z_{24}z_{14}z_{23})^2} + \frac{1}{(z_{14}z_{23}z_{12}z_{34})^2} \right\} + \ldots \tag{4.11}$$

where the dots represent terms of global power smaller than or equal to $-10$ (they disappear in the scaling limit), and $z_{13} = i - k, \ldots$ (real).

The other case, for which $i = (0,0)$, $j = (m,0)$, $k = (0,m)$, $l = (m,m)$ are the vertices of a square of side length $m$, is much more rigid as it depends on a single distance $m$. The result we found for this situation is

$$\text{Prob}[h_i = h_j = h_k = h_l = 1]_{\substack{\text{square} \\ \text{connected}}} = -\frac{3}{8} \frac{[P(1)]^4}{m^8} + \ldots \tag{4.12}$$

Before presenting the results for the other cluster variables of Figure 1, we examine the above correlations for the unit height random variable from the point of view of the conformal field theory which is the most natural candidate, namely the $c = -2$ theory, and its massive extension.

## V. CONFORMAL FIELD THEORY

The $c = -2$ conformal field theory has been studied first in the context of polymers [18], and a bit later served as the simplest example of a logarithmic conformal field theory [19]. Since then it has been extensively examined by many authors [20–26]. Ref. [24] in particular presents a clear and rather complete account of the structure the $c = -2$ theory as a rational conformal field theory. Even if it is considered as the simplest situation where logarithms can occur, it contains many subtle aspects and probably possesses many different and inequivalent realizations. The one which is relevant here is perhaps the most natural one.

The underlying field theory is formulated in terms of a pair of free grassmanian scalars $\theta^\alpha = (\theta, \bar\theta)$ with action

$$S = \frac{1}{2\pi} \int \varepsilon_{\alpha\beta}\, \partial\theta^\alpha\, \bar\partial\theta^\beta = \frac{1}{\pi} \int \partial\theta\, \bar\partial\bar\theta\,, \tag{5.1}$$

where $\varepsilon$ is the canonical symplectic form, $\epsilon_{12} = +1$.

The zero modes of $\theta, \bar\theta$, call them $\xi$ and $\bar\xi$, have been much discussed. Because the action does not depend on them, the expectation value of anything that does not contain $\theta$ and $\bar\theta$ explicitly, but only their derivatives vanishes identically. In particular the partition function itself vanishes, so the correlation functions are normalized by $Z' = \int \mathcal{D}\theta' \mathcal{D}\bar\theta'\, e^{-S}$, where the primed fields exclude the zero modes $\xi$ and $\bar\xi$.

This normalization implies for instance ($\epsilon^{\alpha\beta} = -\epsilon_{\alpha\beta}$)

$$\langle 1 \rangle = 0\,, \qquad \langle \bar\xi\xi \rangle = 1\,, \tag{5.2}$$

$$\langle \theta^\alpha(z)\theta^\beta(w) \rangle = \epsilon^{\alpha\beta}\,, \qquad \langle \theta^\alpha(z)\theta^\beta(w)\bar\xi\xi \rangle = \epsilon^{\alpha\beta} \log |z - w|\,, \tag{5.3}$$

$$\langle \partial\theta^\alpha(z)\, \partial\theta^\beta(w) \rangle = 0\,, \qquad \langle \partial\theta^\alpha(z)\, \partial\theta^\beta(w)\bar\xi\xi \rangle = \frac{\epsilon^{\alpha\beta}}{2(z-w)^2}\,, \tag{5.4}$$

$$\langle \theta^\alpha(z_1)\theta^\beta(z_2)\theta^\gamma(z_3)\theta^\delta(z_4) \rangle = \epsilon^{\alpha\beta}\epsilon^{\gamma\delta} \log|z_{12}z_{34}| - \epsilon^{\alpha\gamma}\epsilon^{\beta\delta} \log|z_{13}z_{24}| + \epsilon^{\alpha\delta}\epsilon^{\beta\gamma} \log|z_{14}z_{23}|\,. \tag{5.5}$$

As far as derivatives of fields are concerned —as will be the case in the ASM, at least at the conformal point—, one can insert the two zero modes in the correlators, as in (5.4), to take care of the integral on constant fields. The functional integral on non–constant fields then yields the usual form for the correlators, obtained from Wick's theorem and the kernel of the laplacian. Equivalently, one can define the functional integral for derivative fields by keeping the zero modes out, or consider the so–called $\eta$–$\xi$ system [18].

The stress–energy tensor components $T = 2 : \partial\theta\, \partial\bar\theta :$ and $\bar T = 2 : \bar\partial\theta\, \bar\partial\bar\theta :$ have OPEs characteristic of a conformal theory with central charge $c = -2$. The fields $\theta$ and $\bar\theta$ are primary fields with conformal dimensions $(0,0)$, while the bosonic composite field $:\theta\bar\theta:$ has the following OPE with $T$

$$T(z) :\theta\bar\theta: (w) = \frac{-1}{2(z-w)^2} + \frac{\partial :\theta\bar\theta: (w)}{z-w} + \ldots \tag{5.6}$$

It shows that the conformal transformations of $:\theta\bar\theta:$ do not close on itself (and its descendants) but also involve the identity and its descendants, which form a conformal module on their own. Thus the identity and $:\theta\bar\theta:$ generate a Virasoro module, that is reducible but not fully reducible. This is a characteristic feature of logarithmic conformal theories [19]. The field $:\theta\bar\theta:$ is called the logarithmic partner of the identity. It is neither a primary field nor a descendant (see below for a field that is primary and descendant without being null).

The fact that there are two fields with zero scaling dimension is the main source of unusual features (and confusing subtleties !), one of them being the existence of two degenerate vacua $|0\rangle$ and $|\xi\bar\xi\rangle$ (there are two more of fermionic nature, $|\xi\rangle$ and $|\bar\xi\rangle$). The above prescription about the insertion of the zero modes can be viewed in the operator formalism as the taking of operator matrix elements between two distinct in–going and out–going vacua.

In conclusion, the theory specified by the action (5.1) is a logarithmic conformal theory with central charge $c=-2$. It contains a non–logarithmic local sector, that retains the central charge value $c=-2$, and in which derivative fields only are considered. Anticipating the analysis to be given below, our results suggest that the ASM scaling fields related to height variables precisely lie in this $c=-2$ non–logarithmic conformal theory.

It should also be noted that either theory, logarithmic or non–logarithmic, contains additional non–local (twisted) sectors. Although they could play an important role in the sandpile models, for the description of other lattice variables than heights, we will not discuss them here, and refer to [24] for further details.

We will also need the off–critical, massive extension of the above conformal theory. It corresponds to a perturbation by the logarithmic partner of the identity

$$S(M) = \frac{1}{\pi}\int \,:\partial\theta\,\bar\partial\bar\theta: + \frac{M^2}{4}\,:\theta\bar\theta:\,. \tag{5.7}$$

The zero mode problem no longer arises in the massive theory, so that one can normalize the correlation functions by the full partition function $Z(M) = \int \mathcal{D}\theta\mathcal{D}\bar\theta\,e^{-S(M)}$. One then obtains

$$\langle\theta(z)\bar\theta(w)\rangle = K_0(M|z-w|), \qquad \langle\theta(z)\theta(w)\rangle = \langle\bar\theta(z)\bar\theta(w)\rangle = 0, \tag{5.8}$$

and for instance

$$\langle\partial\theta(z)\,\bar\partial\bar\theta(0)\rangle = -\frac{M^2}{4}\,\frac{\bar z}{z}\left[2K_0''(M|z|) - K_0(M|z|)\right]. \tag{5.9}$$

On account of $K_0(x) \sim -\log x$ for small arguments, the massless limit of the previous equation exists and reproduces the expression given in (5.4) with the zero modes inserted. This is expected since the effect of the zero mode insertion is formally to change the normalization factor from $Z'$ to $Z$. On the other hand, the same does not apply to the correlations of the fields $\theta^\alpha$ themselves, as the normalizing functional $Z(M)$ goes to zero as $M \to 0$.

As mentioned above, the cluster variables we consider in this article are all related to derivative fields. The previous remark then implies that the off–critical ASM multi–site probabilities have a smooth massless limit, equal to the critical probabilities. The scaling form of the off–critical probabilities will be related to the above massive free theory, while the critical ones will be computable in terms of the non–logarithmic conformal field theory using the insertion prescription.

## VI. SCALING FIELDS FOR CLUSTER VARIABLES

Let us now reconsider the multi–site probabilities for height 1 computed in the Section IV. The 2–site probability suggested that the unit height variable is described by a field with scaling dimension 2, that should be in addition scalar since a unit height variable is rotationally invariant. If one assumes that this field is local in $\theta,\bar\theta$, the only possibilities are $:\partial\theta^\alpha\,\bar\partial\bar\theta^\beta:$, $:\theta\bar\theta\,\partial\theta^\alpha\,\bar\partial\bar\theta^\beta:$ and $M^2:\theta\bar\theta:$. The second set of fields $:\theta\bar\theta\,\partial\theta^\alpha\,\bar\partial\bar\theta^\beta:$ must be excluded, because as explained above, they would produce logarithms in correlation functions, contradicting the observation we made in section II that, in the massless sandpile model ($x=4$), the multi–site probabilities are never logarithmic (at least those one can compute from the Majumdar–Dhar technique, i.e. from finite determinants).

It is not difficult to see that

$$\phi_0 = -P(1)\bigl[\,:\partial\theta\bar\partial\bar\theta + \bar\partial\theta\partial\bar\theta: + \frac{M^2}{2\pi}\,:\theta\bar\theta:\,\bigr] \tag{6.1}$$

is indeed the right combination: its 2–point function is exactly the form given in (4.6), which was obtained by taking the scaling limit of the 2–site probability computed on the lattice.

In order to confirm this identification, the field–theoretic 3–point function of $\phi_0$ can be computed and compared with the lattice result. In the same notations as in Section IV, one finds for an arbitrary arrangement of the insertion points

$$\langle \phi_0(z_1) \phi_0(z_2) \phi_0(z_3) \rangle = -\frac{M^6}{16} \times \left\{ \frac{1}{2} \left( \frac{z_{13} \bar{z}_{23}}{\bar{z}_{13} z_{23}} + \text{c.c.} \right) K_0(12) [2K_0''(13) - K_0(13)] [2K_0''(23) - K_0(23)] + \text{perm.} \right.$$
$$+ \frac{\pi - 2}{\pi^2} \left[ \left( \frac{z_{13} \bar{z}_{23}}{|z_{13}| |z_{23}|} + \text{c.c.} \right) K_0(12) K_0'(13) K_0'(23) + \text{perm.} \right] \quad (6.2)$$
$$+ \frac{1}{\pi} \left[ \left( \frac{z_{12}^2 \bar{z}_{13} \bar{z}_{23}}{|z_{12}|^2 |z_{13}| |z_{23}|} \right) [2K_0''(12) - K_0(12)] K_0'(13) K_0'(23) + \text{perm.} \right] + \left. \frac{\pi^3 - 4}{\pi^3} K_0(12) K_0(13) K_0(23) \right\},$$

where the permutations that must be added are the two exchanges $z_1 \leftrightarrow z_3$ and $z_2 \leftrightarrow z_3$. One easily checks that it reproduces the 3–site probabilities reported in section IV for the two arrangements examined there. The massless limit of (6.2) vanishes, as it clearly follows from (6.1) for $M = 0$, since the 3–point function will necessarily involve a Wick contraction of a $\partial \theta^\alpha$ with some $\bar{\partial} \theta^\beta$.

Finally the 4–point function can be compared. For convenience, we give the field–theoretic result in the massless regime:

$$\langle \phi_0(z_1) \phi_0(z_2) \phi_0(z_3) \phi_0(z_4) \rangle_{M=0} = \frac{P(1)^4}{4|z_{12}|^4 |z_{34}|^4} + \frac{P(1)^4}{4|z_{13}|^4 |z_{24}|^4} + \frac{P(1)^4}{4|z_{14}|^4 |z_{23}|^4}$$
$$- \frac{P(1)^4}{8} \left\{ \frac{1}{(z_{12} z_{34} \bar{z}_{13} \bar{z}_{24})^2} + \frac{1}{(z_{13} z_{24} \bar{z}_{14} \bar{z}_{23})^2} + \frac{1}{(z_{14} z_{23} \bar{z}_{12} \bar{z}_{34})^2} + \text{c.c.} \right\}, \quad (6.3)$$

where only the last term within the curly brackets represents the connected part of the 4–point function. When the four insertions lie on the real axis, it clearly reproduces the lattice result (4.11), and when they are the vertices of a square of side $m$, $z_1 = 0$, $z_2 = m$, $z_3 = im$, $z_4 = (1+i)m$, it reduces to

$$\langle \phi_0(z_1) \phi_0(z_2) \phi_0(z_3) \phi_0(z_4) \rangle_{M=0, \text{ square, connected}} = -\frac{3}{8} \frac{[P(1)]^4}{m^8}, \quad (6.4)$$

and again matches the connected 4–site probability (4.12).

We believe these comparisons provide enough evidence to assert that the unit height random variable of the sandpile model goes over, in the scaling limit, to the field $\phi_0$ defined in (6.1). In the conformal limit, $\phi_0 \sim :\partial \theta \bar{\partial} \bar{\theta} + \bar{\partial} \theta \partial \bar{\theta}: = \partial \bar{\partial} : \theta \bar{\theta}:$ is a primary field with conformal dimensions (1,1), but is also a descendant of $:\theta \bar{\theta}:$.

The rest of this section presents analogous results for the other cluster variables pictured in Figure 1.

We have repeated, for the other thirteen clusters in Figure 1, the same calculations we performed for the unit height variable. More precisely, for each of the cluster variables $S_1$ up to $S_{13}$, we have computed its joint probability with a unit height, namely $\text{Prob}[S_0(0), S_k(i)]$, with $i$ on the principal and on the diagonal axis. From these two probabilities one can write down an Ansatz for the field $\phi_k$ to which the cluster $S_k$ gets identified in the scaling limit. These identifications were subsequently checked to reproduce all 2–site probabilities $\text{Prob}[S_k(0), S_\ell(i)]$, for all pairs $k, \ell = 0, 1, 2, \ldots, 13$, both on the principal and the diagonal axes. In addition, at least one rotated (or mirrored) version of each cluster has been examined, although not systematically (only the correlation with $S_0$ on both axes). The results we found for the rotated clusters are in agreement with the rotations of the fields assigned to the unrotated clusters, so that the field of the rotated cluster is the rotated field. Finally, mixed 3–cluster probabilities involving unit heights and $S_1$ clusters have also been computed. They all confirmed the field identifications.

All calculations have been performed exactly, i.e. not numerically. The 2–cluster probabilities take a form similar to (4.4), where the coefficients are in general complicated rational expressions of $\pi$. Keeping these coefficients in an exact form allows the check of the field identifications to be made in an exact way. For simplicity however, the results presented below are given numerically.

The features of the 2–cluster probabilities are the same as for the unit height variables. We found that all of them scale like $t^2$, with all logarithmic singularities cancelled out. It implies that all cluster variables go in the scaling limit to fields with scaling dimension 2:

$$\lim_{a \to 0} \frac{1}{a^2} \left[ \delta \left( S(i) \right) - P(S) \right] = \phi_S(z), \quad i = \frac{z}{a} \to \infty, \quad t \equiv x - 4 = a^2 M^2 \to 0 \quad \text{with } i\sqrt{t} = Mz. \quad (6.5)$$

This is somewhat surprising as one could have expected the dimension of the scaling fields to increase with the size of the clusters.

All cluster variables we have considered have a scaling limit that corresponds to a field of the following form:

$$\phi(z) = -\left\{ A :\partial\theta\bar{\partial}\bar{\theta} + \bar{\partial}\theta\partial\bar{\theta}: + B_1 :\partial\theta\partial\bar{\theta} + \bar{\partial}\theta\bar{\partial}\bar{\theta}: + iB_2 :\partial\theta\partial\bar{\theta} - \bar{\partial}\theta\bar{\partial}\bar{\theta}: + C\,P(S)\,\frac{M^2}{2\pi} :\theta\bar{\theta}: \right\}. \quad (6.6)$$

The (real) coefficients $A$, $B_1$, $B_2$ and $C$ are given in the Table for each cluster. The factor $P(S)$ in front of the term $:\theta\bar{\theta}:$ is the probability of $S$ evaluated at $x = 4$. Note that the field is not invariant under a rotation of $\frac{\pi}{2}$ as soon as $B_1$ or $B_2$ is non–zero, but is invariant under a rotation of $\pi$ no matter what the coefficients are. So in particular, the scaling limit of the cluster variables does not yield in general conformal fields, but sums of pieces with different tensor structures.

| cluster $S$ | $P(S)$ | $A$ | $B_1$ | $B_2$ | $C$ |
|---|---|---|---|---|---|
| ①  | 0.0736362 | 0.0736362 | 0 | 0 | 1 |
| ②–① | 0.0103411 | 0.0201433 | $-0.00619014$ | 0 | 2 |
| ②–②–① | 0.00141994 | 0.00449027 | $-0.00208908$ | 0 | 3 |
| ①  ②–② | 0.00134477 | 0.00389417 | 0 | $-0.000534524$ | 3 |
| ②–②–②–① | 0.00019246 | 0.000893234 | $-0.000502885$ | 0 | 4 |
| ① ②–②–② | 0.000179829 | 0.000752599 | $-0.000172832$ | $-0.000122784$ | 4 |
| ②–① ②–② | 0.000173323 | 0.000695941 | $-0.00012949$ | $-0.000153467$ | 4 |
| ① ② ②–② | 0.000179829 | 0.000752599 | 0.000172832 | $-0.000122784$ | 4 |
| ① ②–③ ① | 0.000173106 | 0.000692147 | 0.000135489 | 0 | 4 |
| ① ②–③–① | 0.000173106 | 0.000692147 | $-0.000135489$ | 0 | 4 |
| $S_{10}$ | 0.0000572863 | 0.000255127 | $-0.0000470539$ | $-0.0000259802$ | 5 |
| $S_{11}$ | 0.00000731457 | 0.000042272 | $-0.0000130978$ | $-0.0000071978$ | 6 |
| $S_{12} = $ ①–◯–① | 0.00496687 | 0.00969315 | $-0.00129393$ | 0 | 2 |
| $S_{13} = $ ◯–① ①–◯ | 0.00404859 | 0.010213 | 0 | $-0.00312009$ | 2 |

TABLE I. For each cluster in Figure 1, the Table gives the values of the parameters $A$, $B$ and $C$ specifying the field that describes the scaling behaviour of the given cluster (see Eq. (6.6)). Note in particular that the coefficient $C$ is equal to the size of the cluster.

As far as numerical values are concerned, the last column of the Table is particularly striking: all entries are integers,

simply equal to the size of the cluster. This makes the coefficient of the $:\theta\bar{\theta}:$ terms particularly simple and apparently regular. The reason for this is unclear.

The other numbers mentioned in the Table are not in themselves particularly interesting. As mentionned above, all these numbers are complicated expressions. For instance the first three numbers on the line corresponding to $S_9$ (the last cluster of size 4) are in fact equal to

$$\text{Prob}(S_9) = \frac{2621440}{27\,\pi^7} - \frac{21389312}{81\,\pi^6} + \frac{24279040}{81\,\pi^5} - \frac{14968672}{81\,\pi^4} + \frac{1809776}{27\,\pi^3} - \frac{258037}{18\,\pi^2} + \frac{10061}{6\,\pi} - \frac{663}{8}, \tag{6.7}$$

$$A = \left(\frac{3\pi - 8}{\pi^2}\right)\left(\frac{655360}{27\,\pi^5} - \frac{3389440}{81\,\pi^4} + \frac{2259952}{81\,\pi^3} - \frac{81566}{9\,\pi^2} + \frac{5765}{4\,\pi} - \frac{8647}{96}\right), \tag{6.8}$$

$$B_1 = \left(\frac{3\pi - 8}{\pi^2}\right)\left(\frac{305152}{81\,\pi^4} - \frac{359056}{81\,\pi^3} + \frac{17554}{9\,\pi^2} - \frac{13693}{36\,\pi} + \frac{2663}{96}\right). \tag{6.9}$$

A gross feature of the Table is that the (non–zero) numbers are roughly constant for all clusters of the same size, namely the probabilities and the coefficients do not change much with the shape of the clusters, but depend essentially on their size only. Roughly speaking, these numbers (except $C$) get divided by 10 when the size increases by one.

The zeroes in the Table or the equality (up to signs) of coefficients can be understood from the transformations of the clusters and the corresponding fields under the symmetry group of the lattice. One easily sees that the field $\phi$ in (6.6) changes under rotations and reflections according to the following rules:

$$(A, B_1, B_2, C) \longrightarrow \begin{cases} (A, -B_1, -B_2, C) & \text{under a } \frac{\pi}{2}\text{–rotation,} \\ (A, B_1, -B_2, C) & \text{under a } x\text{– or } y\text{–reflection.} \end{cases} \tag{6.10}$$

By convention, all clusters are assumed to be anchored to their lower left site. The rotations are performed about an axis passing through that site.

First of all, the only one to have $B_1 = B_2 = 0$ is the unit height. Indeed it is the only cluster that preserves its shape under rotations and reflections, and so one can expect the corresponding field to be a scalar under (continuous) rotations and reflections.

There are clusters whose fields have $B_2 = 0$, and they are precisely those clusters which are invariant under a reflection through the horizontal axis. The same can be said of the rotated clusters for a reflection through the vertical axis. That $S_9$ has a coefficient $B_2 = 0$ can be understood along the same lines, although it is not manisfestly invariant under reflections. An $x$–reflection of $S_9$ followed by a rotation by $\pi$ and a translation of two lattice sites bring it to itself, except that a height 2 and a height 1 have been swapped. However, the assignment of heights within a cluster is irrelevant in the actual computation: the modified ASM is defined in terms of certains bonds being removed. Since each site whose height is being constrained looses three out of his four bonds, the actual height assignment is irrelevant. In effect, the set $M_S$ which includes all the sites affected by the modifications and the modification matrix $B$ itself can be chosen (have been chosen) invariant under a $y$–reflection.

In the same way, one sees that $S_5$ and $S_7$ have equal coefficients, up to signs. As represented in the Table and in Figure 1, they are related by rotation of $\frac{\pi}{2}$ and an $x$–reflection, with the consequence that their $B_1$ coefficients are opposite but the $B_2$ are equal. The same can be said of $S_8$ and $S_9$, with the same remark as above regarding the locations of the height values within the clusters.

From these remarks, one easily finds the fields corresponding to different orientations of a cluster. The cluster $S_6$ for instance comes in eight different orientations (all anchored to the same site). All of them have the same coefficient $A \sim 0.000695941$ and $C = 4$, whereas pairs of clusters have coefficients $(B_1, B_2)$, or $(-B_1, B_2)$, or $(B_1, -B_2)$ or $(-B_1, -B_2)$. As a consequence, the sum over the corresponding eight fields reduces to a projection onto the scalar part, and involves the $A$ and $C$ terms only.

In a sense, the fact that the fields reflect so well the geometric symmetries of the clusters is surprising. As discussed at length in Section II, the actual calculations are based on adequate modifications of the original ASM on a set we called $M_S$, which not only contains sites belonging to the cluster itself, but also sites in its close neighbourhood. Thus each cluster drags with itself an invisible shadow, made of the sites in the set $M_S \setminus S$. The shadow is a computational artefact, but is nevertheless crucial. Moreover it usually breaks or alters the geometric symmetries of the cluster it goes with. The insertion of a height 1 for instance, somewhere in the lattice, really requires to consider a 4–cluster pictured in Section II. Here the shadow consists of three neighbours of the central site, and clearly breaks the rotational invariance.

We will conclude this section by observing that the height $h$ variables, for $h$ bigger than 4, can be handled in the massive ASM exactly like the unit height variables, even more simply. The reason is that a height equal to $5, 6, \ldots, x$ can never be in a forbidden subconfiguration, so that the set of recurrent configurations containing a height equal to $h > 4$ at some site $i$ is equal to the set of recurrent configurations on the lattice with $i$ removed. Therefore the modifications needed to freeze the height of a site to $h > 4$ must simply reduce the threshold at that site to 1, and cut it off from the rest of the lattice. This can be implemented by the following matrix

$$B = \begin{pmatrix} 1-x & 1 & 1 & 1 & 1 \\ 1 & 0 & 0 & 0 & 0 \\ 1 & 0 & 0 & 0 & 0 \\ 1 & 0 & 0 & 0 & 0 \\ 1 & 0 & 0 & 0 & 0 \end{pmatrix}. \tag{6.11}$$

The corresponding probability is simply given by $\text{Prob}[h_i = h > 4] = G(0,0)$, and is logarithmically divergent at $x = 4$. For that reason, one considers instead the probability that $h_i$ exceeds 4:

$$\text{Prob}[h_i > 4] = (x-4)\, G(0,0) = \frac{2(x-4)}{\pi x}\, K\left(\frac{4}{x}\right), \tag{6.12}$$

which goes to 0 when $x \to 4$. (The matrix $B$ corresponding to this has $-4$ as first diagonal entry, rather than $1-x$.) $K$ is a complete elliptic function (see Appendix A).

As for the above clusters, one can compute the correlations of this random variable $\delta(h_i > 4)$ with itself or with the other clusters, and see what field–theoretic description it has in the scaling limit.

Again the result is simple. The lattice calculation of its own correlation yields

$$\text{Prob}[h_0 > 4,\, h_i > 4] - \text{Prob}[h_0 > 4]^2 = -\frac{t^2}{4\pi^2}\, K_0^2(\sqrt{t}|i|), \tag{6.13}$$

which suggests the scaling limit

$$\delta(h_i > 4) - \langle \delta(h_i > 4)\rangle \quad \overset{\text{scaling}}{\longrightarrow} \quad \phi = \frac{M^2}{2\pi}\, {:}\theta\bar\theta{:}\,. \tag{6.14}$$

Correlations with the other cluster variables confirm this limit. It nicely fits the expectation that the field should vanish at the critical point.

## VII. THE HEIGHT TWO VARIABLE

We have so far focused on the class of weakly allowed cluster variables, whose correlations can be handled by the technique developed in [8], and in turn computed from a finite determinant. The authors point out in that article that non weakly allowed cluster variables can in fact be viewed as infinite series of weakly allowed clusters. It dramatically complicates their treatment, since a correlation involving a single non weakly allowed cluster requires to compute an infinite number of correlations of weakly allowed clusters, of finite but unbounded size.

In this section, we address the question of the field assignment for the height two variable, in the light of the results of the previous sections. We will consider the height two variable, both from the perturbative point of view that we have just summarized, and from the conformal point of view.

That a height two variable can be treated as an infinite sum of weakly allowed cluster variables can be seen as follows [8]. Consider the set of recurrent configurations $\mathcal{C}$ with a height 2, at the origin say. That set can be divided up into two disjoint subsets according to whether the configurations remain recurrent when the 2 is replaced by a 1, or become transient upon that replacement.

The number of those that remain recurrent is the same as the number of recurrent configurations which have a height 1 at the origin, because vice–versa, a recurrent configuration with a 1 remains recurrent if the 1 is replaced by a 2. So the contribution to $P(2) \equiv \text{Prob}[h_0 = 2]$ from this first subset is exactly equal to $P(1)$.

For those configurations which become transient, it must be that the 2 belongs to a weakly allowed cluster. This weakly allowed cluster can be of various size and shape, and a straight enumeration according to their size leads directly to the clusters of Figure 1 (except the first one and the last two) and their various orientations. In this way, the second subset is itself divided into a infinite number of disjoint subsets, according to which weakly allowed cluster $S$ the height 2 at the origin is part of. The subset labelled by $S$ (fixed size, shape and orientation) contributes $P(2)$ a term equal to $P(S)$.

Putting all together, one obtains, observing that the number $P(S)$ does not depend on the orientation of $S$, the formula

$$P(2) = P(1) + \sum_{\text{w.a.c. } S} P(S) = P(1) + 4P(S_1) + 4P(S_2) + 8P(S_3) + 4P(S_4) + 8P(S_5) + \ldots \tag{7.1}$$

where the summation is over the weakly allowed clusters which are 'anchored' to a height 2. As pointed out in [8], the convergence is very slow. From Table 1, the terms up to $S_9$ furnish the lower bound $P(2) \geq 0.13855$, well below the exact value $P(2) \sim 0.1739$ [12].

The argument recalled above leading to the perturbative formula for $P(2)$ works similarly for any correlation. The result can be expressed as an identity between random variables,

$$\delta(h_i - 2) = \delta(h_i - 1) + \sum_{\text{w.a.c. } S} \delta(S(i)). \tag{7.2}$$

Modulo the issue of convergence, this identity is valid when inserted in expectation values.

The results of the previous section suggest that all random variables on the r.h.s. have the same scaling form, given by the field in (6.6). Assuming this at all orders, and taking the scaling limit of the previous identity lead to a scaling field for the height two of the same form as the scaling field for the height one, namely

$$\delta(h_i - 2) \stackrel{\text{scaling}}{\longrightarrow} \alpha : \partial\theta\bar\partial\bar\theta + \bar\partial\theta\partial\bar\theta : + \beta\, M^2 :\theta\bar\theta: . \tag{7.3}$$

This follows from the observation we made earlier that the other terms $\partial\theta\partial\bar\theta \pm \bar\partial\theta\bar\partial\bar\theta$ change sign under a rotation by $\frac{\pi}{2}$. The sum over the orientations of a cluster make these terms cancel against each other, leaving a scalar field, as it should.

The natural conclusion one could draw from this is that, at the critical point, the heights one and two scale the same way and in fact go over, in the scaling limit, to the same —up to normalization— primary field of conformal dimensions (1,1). This is the first direct though teneous evidence in favour of such a statement, which has in fact been made in [11], based on an extrapolation to the bulk of a similar statement on the corresponding boundary variables, itself relying on the boundary 2–point functions. As plausible and likely as it may be, the extrapolation remains uncontrolled, as there are well–known examples of lattice observables that go to different fields, depending on whether they lie on a boundary or in the bulk. Thus neither argument is convincing, but both point to the same field assignment for the height two variable (and probably similarly for the heights three and four).

This seems reasonable and likely. It is therefore surprising to observe that it does not appear to be consistent with a naive interpretation of the operator product expansions (OPE). To simplify, we consider the critical point, and the corresponding conformal field theory.

The two lattice variables, a height one and a height two, can be taken far apart, and subsequently brought closer to each other, until they occupy neighbouring sites, then forming the cluster variable we called $S_1$. In the field–theoretic picture, this amounts to taking the two corresponding fields closer and closer to each other, until they become coincident, at which point they form a new composite field. The information about what composite fields a pair of fields can form when they come close to each other and asymptotically coincident, is contained in their operator product expansion (OPE).

Thus it seems natural to expect that the field assigned to the cluster variable ②–① be in the OPE of the field corresponding to the height one with the field corresponding to the height two. If one assumes, as argued above, that the heights one and two scale to the same field, the required OPE is simply

$$:\partial\theta\bar\partial\bar\theta + \bar\partial\theta\partial\bar\theta: (z) \; :\partial\theta\bar\partial\bar\theta + \bar\partial\theta\partial\bar\theta: (w) = -\frac{1}{2|z-w|^4} + \frac{:\bar\partial\theta\bar\partial\bar\theta:(w)}{(z-w)^2} + \frac{:\partial\theta\partial\bar\theta:(w)}{(\bar z - \bar w)^2} + \text{less singular}, \tag{7.4}$$

where, from dimensional analysis, the less singular terms involve fields of scale dimension strictly larger than 2. One sees from (7.4) that the only fields with scale dimension 2 which can be formed in the fusion of a height one with a height two are the non–scalar parts of the field making the cluster ②–①. The scalar part of it, $:\partial\theta\bar\partial\bar\theta + \bar\partial\theta\partial\bar\theta:$, is missing. (Note that it must be so, since otherwise the unit height variables, represented by that scalar field, would have a non–zero (connected) 3–point function.)

One may observe that the only dimension 2 scalar fields whose fusion produces all field components of ②–① are logarithmic fields, like $:\theta\bar\theta(\partial\theta\bar\partial\bar\theta + \bar\partial\theta\partial\bar\theta):$. The change for this logarithmic scalar field has however heavy consequences as correlations involving heights two would automatically contain logarithmic functions of the separation distances, in addition to the usual rational functions.

Note that for exactly the same reasons, one could question the field assignment of the height one variable itself, despite the fact that the field $\phi_0$ has successfully passed so many tests. Although one cannot bring two heights one side by side, one can bring them fairly close to each other, like in the last two clusters of Figure 1 (or Table 1), in fact close enough so as not to loose the OPE argument. But then the fields associated to the two clusters $S_{12}$ and $S_{13}$ must be contained in the fusion of two heights one, i.e. in the fusion (7.4), which we know is not the case.

Perhaps sandpile models are so special that one would reject the fusion altogether, on the basis that height variables have hair ! Because a particular height imposes restrictions on what can stand close to it. For example a height 1 forces all its neighbours to be higher or equal to 2, and a height 2 does not allow two of its neighbours to have a height 1. This might explain the unconsistency noticed above, but at the same time it denies the very possibility of a field assignment. We believe that this issue should be clarified.

## VIII. CONCLUSION

The power of conformal field theory could bring a much better understanding of the sandpile model, if some of its observables could be identified with conformal fields. This is a non–trivial task even for the height variables, which are probably the easiest variables to account for in a field–theoretic setting. In addition, and in order to strengthen the connection with a field theory, the neighbourhood of the critical point should be investigated. In this article, we have taken the first steps towards a systematic study of this relationship, at and off–criticality.

The off–critical extension of the sandpile that we considered is defined by allowing dissipation, i.e. loss of sand each time a site topples. The dissipation rate is controlled by a parameter $t \geq 0$ and corresponds to a relevant perturbation of the usual abelian undirected sandpile model.

We have examined multi–site probabilities for the simplest local cluster variables in the off–critical sandpile model. By explicit calculations, we have shown that their scaling form can be fully reproduced by a free field theory of massive grassmanian scalars. In the massless, critical limit, this theory is a logarithmic conformal field theory with central charge $c = -2$. The local fields assigned to the various cluster variables however all belong to a non–logarithmic bosonic sector. The massive regime, with a mass $M \sim \sqrt{t}$ directly related to the perturbing parameter, corresponds to a thermal perturbation of the conformal theory, i.e. a mass term specified by a logarithmic field.

We have determined the field assignment for the fourteen cluster variables pictured in Figure 1, and checked their consistency against the correlation functions. On the other hand, at the critical point, we have noted a disagreement between these assignments and the naive fusion rules of the conformal theory.

We do not claim that all features of the sandpile models will be comprehensible within a field theory, but some of them definitely are. In this respect, other issues than the height variables can be raised: boundary phenomena against boundary conformal field theory, the question of the modular non–invariance on a torus (with leaking sites), ... Also the relevance and the role of logarithmic fields and twist fields in the $c = -2$ logarithmic conformal field theory must be further examined.

## ACKNOWLEDGMENTS

P.R. heartily thanks Michael Flohr for many stimulating discussions and for sharing his own insight into the subtleties of logarithmic conformal field theories. Useful discussions with Deepak Dhar about the draft of this article are also gratefully acknowledged.

## APPENDIX A: GREEN FUNCTIONS

In this appendix, we collect a number of expressions we have used for the computations of correlations in the sandpile model.

The central object here is the Green function $G$ of the massive discrete laplacian on $\mathbb{Z}^2$, that is the solution of the Poisson equation $\Delta G = 1$, with $\Delta$ being the finite difference operator given in (3.1). The solution is easily obtained by Fourier transform

$$G(m,n) = G((m',n'),(m+m',n+n')) = \iint_0^{2\pi} \frac{\mathrm{d}^2 k}{4\pi^2} \frac{\mathrm{e}^{ik_1 m + ik_2 n}}{x - 2\cos k_1 - 2\cos k_2}, \qquad (m,n),(m',n') \in \mathbb{Z}^2. \tag{A1}$$

As explained in the text, values of $G$ are needed at points $x$ which are either close to the origin, or else very far from the origin, and in this last case, we have restricted ourselves to points close to a principal or a diagonal axis. We treat these three cases in turn.

## 1. The Green function at points close to the origin

By using the invariance of $G$ under the reflection symmetries of the lattice and its defining equation $\Delta G = 1$, the Green function can be given everywhere in terms of its values on a diagonal. By a suitable change of variables and one integration [27], the diagonal values can be recast into

$$G(m,m) = \frac{(-1)^m}{\pi x} \int_0^\pi dt \, \frac{\cos 2mt}{\sqrt{1 - \frac{16}{x^2} \sin^2 t}}. \tag{A2}$$

This can be resolved in terms of the complete elliptic functions [28]

$$K(p) = \int_0^{\pi/2} dt \, \frac{1}{\sqrt{1 - p^2 \sin^2 t}} = \left[ 1 + \frac{q^2}{4} + \frac{9q^4}{64} + \ldots \right] \log\left(\frac{4}{q}\right) - \left[ \frac{q^2}{4} + \frac{21q^4}{128} + \ldots \right], \tag{A3}$$

$$E(p) = \int_0^{\pi/2} dt \, \sqrt{1 - p^2 \sin^2 t} = \left[ 1 - \frac{q^2}{4} - \frac{13q^4}{64} - \ldots \right] + \left[ \frac{q^2}{2} + \frac{3q^4}{16} + \ldots \right] \log\left(\frac{4}{q}\right), \tag{A4}$$

where $q = \sqrt{1 - p^2}$, and where the expansions are given for $p \lesssim 1$ close to 1.

In terms of our perturbing parameter $t = x - 4$, one finds for instance

$$G(0,0) = \frac{2}{\pi x} K\left(\frac{4}{x}\right), \tag{A5}$$

$$G(1,1) = \frac{1}{4\pi x} \left\{ (x^2 - 8) K\left(\frac{4}{x}\right) - x^2 E\left(\frac{4}{x}\right) \right\}, \tag{A6}$$

$$G(2,2) = \frac{1}{24\pi x} \left\{ (x^4 - 16 x^2 + 48) K\left(\frac{4}{x}\right) - x^2 (x^2 - 8) E\left(\frac{4}{x}\right) \right\}, \tag{A7}$$

$$G(3,3) = \frac{1}{120\pi x} \left\{ (x^6 - 24 x^4 + 158 x^2 - 240) K\left(\frac{4}{x}\right) - x^2 (x^4 - 16 x^2 + 46) E\left(\frac{4}{x}\right) \right\}, \tag{A8}$$

which can then be expanded around $x = 4$ by using (A3) and (A4). They all have the same logarithmic singularity at $x = 4$ as $G(0,0)$, so that the differences $G(m,n) - G(0,0)$ remain finite when $x \to 4$. In particular, the critical limit of the subtracted diagonal Green function is simply [27]

$$\lim_{x \to 4} [G(m,m) - G(0,0)] = -\frac{1}{\pi} \sum_{k=1}^m \frac{1}{2k-1}. \tag{A9}$$

## 2. The Green function on the far diagonal

For $m$ large, the use of elliptic functions is impractical to extract the asymptotic behaviour in $m$. Making the change of variables $z = e^{it}$, the formula (A2) becomes a integral over a contour that can be deformed to enclose the cut lying in between the two roots $\pm u$ of the denominator, with $u = \frac{x}{4} - \sqrt{\frac{x^2}{16} - 1}$. This yields

$$G(m,m) = -\frac{1}{2\pi} \int_{-u}^{u} dz \, \frac{z^{2m}}{\sqrt{(z^2 - u^2)(z^2 - \frac{1}{u^2})}}. \tag{A10}$$

The asymptotic expansion of this kind of integral was studied in [29], from which one finds, using their notations

$$G(m,m) = \sqrt{\frac{u}{x(1-u^2)}} \frac{1}{\sqrt{2\pi m}} u^{2m} \left\{ 1 + \frac{\tilde{A}_{1>}}{8m} + \frac{3}{64} \frac{\tilde{A}_{2>} - \frac{5}{6}}{m^2} + \frac{15}{512} \frac{\tilde{A}_{3>} - \frac{7}{6}\tilde{A}_{1>}}{m^3} + \ldots \right\}, \tag{A11}$$

where the coefficients $\tilde{A}_{n>}$ are defined from the generating function

$$\tilde{A}_>(z) = \frac{1}{\sqrt{(1 + \frac{1-u^2}{1+u^2} z)(1 + \frac{1+u^2}{1-u^2} z)}} = \sum_{n=0}^\infty \tilde{A}_{n>} z^n, \tag{A12}$$

and are thus themselves infinite (Laurent) series in $u^2$, hence in $\sqrt{t} = \sqrt{x-4}$. It is not difficult to show that these coefficients start off like

$$\tilde{A}_{n>} = (-1)^n \frac{(2n-1)!!}{(2n)!!} \left(\frac{2}{t}\right)^{n/2} + \mathcal{O}(t^{-\frac{n}{2}+1}), \tag{A13}$$

with the consequence that the $m^{-n}$ term in (A11) takes the form

$$\frac{(2n-1)!!}{8^n} \frac{\tilde{A}_{n>} + \ldots}{m^n} = \frac{(-1)^n}{n!} \left(\frac{(2n-1)!!}{2^n}\right)^2 \frac{1}{(2\sqrt{2t}\,m)^n} \left[1 + \text{series in } t\right], \tag{A14}$$

that is, a first term which has the scaling form times corrections in $t$, independent of the distance $m$.

By combining the previous expansion with that of the prefactor of (A11),

$$\sqrt{\frac{u}{x(1-u^2)}} \frac{1}{\sqrt{2\pi m}} u^{2m} = \left(\frac{1}{8\pi\sqrt{2t}\,m}\right)^{\frac{1}{2}} e^{-m\sqrt{2t}+m\sqrt{2t^3}/48+\cdots} \left[1 + \text{series in } t\right], \tag{A15}$$

one eventually finds that the Green function can be written as

$$G(m,m) = \left\{ D_0(m\sqrt{2t}) + t\,D_2(m\sqrt{2t}) + t^2\,D_4(m\sqrt{2t}) + \ldots \right\} e^{m\sqrt{2t^3}/48+\cdots}, \tag{A16}$$

where all functions $D_i$ depend on the single scaling variable $m\sqrt{2t}$ (the square root of 2 has to be included, since the distance from the origin is $\sqrt{2}m$). Moreover, from (A14), the first function $D_0$ is explicitely given as

$$D_0(z) = \frac{1}{\sqrt{8\pi z}} e^{-z} \sum_{n=0}^{\infty} \frac{(-1)^n}{n!} \left(\frac{(2n-1)!!}{2^n}\right)^2 \frac{1}{(2z)^n} = \frac{1}{2\pi} K_0(z), \tag{A17}$$

a modified Bessel function. This is to be expected and confirms that the scaling limit of the Green function is indeed equal to $\frac{1}{2\pi} K_0(Mr)$, the propagator of a massive scalar.

For calculations in the ASM model, one still needs the Green functions at points close to the diagonal. The Poisson equation is not sufficient, because it would require the knowledge of the Green function all the way down to the horizontal axis, but a simple Ansatz similar to (A16) leads to the following expressions, valid for $0 \leq k \ll m$:

$$\begin{aligned} G(m,m+k) &= \left\{ D_0(z) + k\,D_0'(z)\sqrt{\frac{t}{2}} + \left[D_2(z) + \frac{k^2}{4} D_0(z)\right] t + \right. \\ &\quad \left. \left[\frac{k}{96} D_0(z) + \frac{k^3}{8} D_0'(z) - \frac{k^3}{12} D_0'''(z) + \frac{k}{2} D_2'(z)\right] \sqrt{2t^3} + \ldots \right\} e^{zt/48+\cdots}, \end{aligned} \tag{A18}$$

where $z = m\sqrt{2t}$ is the scaled distance. At the order where all the calculations have been performed, the terms shown in the previous expression are all that is needed.

The critical limit of the above expansions is more conveniently computed from (A9) by using the asymptotic expansion of the $\psi$–function [28], or from the integral (A2). The result is

$$\lim_{x \to 4} \left[G(m,m) - G(0,0)\right] = -\frac{1}{2\pi} \log m - \frac{1}{\pi}\left(\frac{\gamma}{2} + \log 2\right) - \frac{1}{48\,\pi\,m^2} + \frac{7}{1920\,\pi\,m^4} - \frac{31}{16128\,\pi\,m^6} + \ldots \tag{A19}$$

with $\gamma = 0.57721\ldots$ the Euler constant.

### 3. The Green function on the far principal axis

The calculations can be repeated on a principal axis. The integration of (A1) over $k_2$ followed by the change of variable $z = e^{ik_1}$ gives $G(m,0)$ as a contour integral over the unit circle. It can again be deformed to encircle the branch cut joining the two roots $v < u$ of the denominator which lie inside the unit circle, yielding

$$G(m,0) = \frac{1}{4\pi} \int_0^{2\pi} dk_1 \, \frac{e^{ik_1 m}}{\sqrt{\left(\frac{x}{2} - \cos k_1\right)^2 - 1}} = \frac{1}{i\pi} \int_v^u dz \, \frac{z^m}{\sqrt{(z-u)(z-\frac{1}{u})(z-v)(z-\frac{1}{v})}}, \tag{A20}$$

with $u < 1$ and $v < 1$ the two roots of $z^2 - (x-2)z + 1$ and $z^2 - (x+2)z + 1$ respectively, that is,

$$u = \frac{1}{2}[x - 2 - \sqrt{x(x-4)}], \qquad v = \frac{1}{2}[x + 2 - \sqrt{x(x+4)}]. \tag{A21}$$

The asymptotic behaviour of this integral for large $m$ can be found again in [29], with the result

$$G(m,0) = \sqrt{\frac{u}{4\pi(1-u^2)}} \frac{u^m}{\sqrt{m}} \left\{ 1 + \frac{A_{1>}}{4m} + \frac{3}{16} \frac{A_{2>} - \frac{5}{6}}{m^2} + \frac{15}{64} \frac{A_{3>} - \frac{7}{6} A_{1>}}{m^3} + \ldots \right\}. \tag{A22}$$

The series within the curly brackets is similar to that of the previous subsection, with however $\frac{m}{2}$ substituted for $m$, and with the coefficients $A_{n>}$ as defined in [29], namely by (the coefficients $\tilde{A}_{n>}$ used above correspond to the present $A_{n>}$ upon the identification $v = -u$)

$$A_>(z) = \frac{1}{\sqrt{(1 + \frac{1+uv}{1-uv}z)(1 - \frac{1+v/u}{1-v/u}z)(1 + \frac{1+u^2}{1-u^2}z)}} = \sum_{n=0}^{\infty} A_{n>} z^n, \tag{A23}$$

The usual expansions around $x = 4$ now yields

$$G(m,0) = \left\{ P_0(m\sqrt{t}) + t\, P_2(m\sqrt{t}) + t^2\, P_4(m\sqrt{t}) + \ldots \right\} e^{m\sqrt{t^3}/24 + \ldots}, \tag{A24}$$

with $P_0(z) = \frac{1}{2\pi} K_0(z)$ as before.

For points close to the horizontal axis, one finds from Poisson equation the expansions for $k \ll m$

$$G(m,k) = \left\{ P_0(z) + \left[ P_2(z) + \frac{k^2}{2} \left( P_0(z) - P_0''(z) \right) \right] t + \ldots \right\} e^{zt/24 + \ldots}, \tag{A25}$$

with $z = m\sqrt{t}$ the scaled distance. The ASM calculations also need the values of $G(m \pm \ell, k)$ for small $\ell$, and those can easily be obtained by expanding the previous result, yielding a Taylor series in $\sqrt{t}$.

The critical asymptotic expansion of $G(m,0)$ can also be computed from (A20). One has

$$[G(m,0) - G(m+1,0)]\Big|_{x=4} = \frac{1}{4\pi} \int_0^{2\pi} dk_1\, e^{ik_1 m} F(k_1), \tag{A26}$$

where $F(x) = \frac{1 - e^{ix}}{\sqrt{(2 - \cos x)^2 - 1}}$. A repeated use of integration by parts then leads to

$$[G(m,0) - G(m+1,0)]\Big|_{x=4} = -\frac{1}{4\pi} \sum_{k \geq 1} \frac{(-1)^k}{(im)^k} \left[ d_x^k F \right]_0^{2\pi} = \frac{1}{2\pi} \left\{ \frac{1}{m} - \frac{1}{2m^2} + \frac{1}{2m^3} - \frac{1}{2m^4} + \ldots \right\}, \tag{A27}$$

from which one deduces the subtracted Green function itself as

$$\lim_{x \to 4} [G(m,0) - G(0,0)] = -\frac{1}{2\pi} \log m - \frac{1}{\pi} (\tfrac{\gamma}{2} + \tfrac{3}{4} \log 2) + \frac{1}{24\,\pi\,m^2} + \frac{43}{480\,\pi\,m^4} + \frac{949}{2016\,\pi\,m^6} + \ldots \tag{A28}$$

### APPENDIX B: ABOUT THE SINK SITE

The evacuation of sand is a crucial ingredient to the self–organized criticality of the sandpile models. In order for its dynamics to be well–defined —any unstable configuration relaxes to a stable one—, each site should be pathwise connected to a sink, where goes the sand that falls off the pile. The sink is usually omitted in all discussions, perhaps because in the ordinary ASM, only the boundary sites are connected to the sink, and that the large volume limit takes them to infinity. In the massive ASM however, each site is connected to the sink. One might thus worry about a possible role of it in actual computations.

We show here that the sink has in fact no effect at all and can be omitted completely, be it in the usual or massive ASM. The argument is simple and worth being made explicit.

The discrete dynamics of the ASM recalled in the Introduction uses a toppling matrix that ignores the sink site s. To include it, one simply defines an extended toppling matrix $\Delta_e$ by adding to $\Delta$ a row and a column,

$$\Delta_e = \begin{pmatrix} 1 & 0 \\ V & \Delta \end{pmatrix}. \tag{B1}$$

The diagonal entry $(\Delta_e)_{s,s}$ is set equal to 1, in order to freeze the height of the sink site. The rest of the first row is equal to 0, since the sink has no connection to the sites of the pile. On the other hand, the first column is not zero: $V_{i,s} = -n_i$ if $n_i$ grains of sand fall off the pile when site $i$ topples. The number $n_i$ is equal to $n_i = -\sum_j \Delta_{i,j}$, so that all row sums of $\Delta_e$, except the first one, are zero. In the usual ASM, $V_{i,s}$ is non–zero for the boundary sites only, whereas in the massive model, all components are non–zero, with $V_{i,s} = 4 - x$ for all bulk sites. The formula for the number of recurrent configurations remains valid, with obviously the same result, $\det \Delta_e = \det \Delta$.

The same method for computing probabilities and correlations of weakly allowed clusters works as before. One uses an extended $B$ matrix specifying the way the ASM needs be modified

$$B_e = \begin{pmatrix} 0 & 0 \\ W & B \end{pmatrix}. \tag{B2}$$

The first row is clearly always zero, but the first column can be non–zero, depending on the modifications. For those called the "least economical" ones in Section II, in which one cuts the cluster off the rest of the lattice, each site $i$ of the cluster is left with a sole connection to the sink, so one sets $W_i = 5 - x$.

The probability of a cluster variable $S$ is given by the usual formula, which as before reduces to a finite determinant

$$\text{Prob}(S) = \frac{\det(\Delta_e + B_e)}{\det \Delta_e} = \det(\mathbb{I} + \Delta_e^{-1} B_e) = \det(\mathbb{I} + \Delta_e^{-1} B_e)\Big|_{M_S \cup \{s\}}. \tag{B3}$$

However, the restriction to $M_S \cup \{s\}$ of $\Delta_e^{-1} B_e$ is particularly simple,

$$\Delta_e^{-1} B_e = \begin{pmatrix} 1 & 0 \\ -\Delta^{-1} V & \Delta^{-1} \end{pmatrix} \begin{pmatrix} 0 & 0 \\ W & B \end{pmatrix} = \begin{pmatrix} 0 & 0 \\ \Delta^{-1} W & \Delta^{-1} B \end{pmatrix}, \tag{B4}$$

and manifestly leads to the usual result, with no sink,

$$\text{Prob}(S) = \det(\mathbb{I} + \Delta^{-1} B)\Big|_{M_S}. \tag{B5}$$

[1] P. Bak, C. Tang and K. Wiesenfeld, Phys. Rev. Lett. **59** (1987) 381.
[2] P. Bak, *How Nature works – the science of self–organized criticality*, Oxford University Press, Oxford 1997.
[3] H. J. Jensen, *Self–Organized Criticality*, Cambridge University Press, Cambridge 1998.
[4] D. Dhar, *The Abelian Sandpile and Related Models*, cond-mat/9808047.
[5] D. Dhar, *Studying Self–Organized Criticality with Exactly Solved Models*, cond-mat/9909009.
[6] D. Dhar, Phys. Rev. Lett. **64** (1990) 1613.
[7] S.N. Majumdar and D. Dhar, Physica A**185** (1992) 129.
[8] S.N. Majumdar and D. Dhar, J. Phys. A: Math. Gen. **24** (1991) L357.
[9] E.V. Ivashkevich and V.B. Priezzhev, Physica A**254** (1998) 97.
[10] J.G. Brankov, E.V. Ivashkevich and V.B. Priezzhev, J. Phys. I France **3** (1993) 1729.
[11] E.V. Ivashkevich, J. Phys. A: Math. Gen. **27** (1994) 3643.
[12] V.B. Priezzhev, J. Stat. Phys. **74** (1994) 955.
[13] S.N. Majumdar and D. Dhar, J. Phys. A: Math. Gen. **23** (1990) 4333.
[14] C.J. Pérez, Á. Corral, A. Díaz-Guilera, K. Christensen and A. Arenas, Int. J. Mod. Phys. B **10** (1996) 1111.
[15] S.S. Manna, L.B. Kiss and J. Kertész, J. Stat. Phys. **61** (1990) 923.
[16] T. Tsuchiya and M. Katori, Phys. Rev. E **61** (2000) 1183.
[17] A. Vázquez, Phys. Rev. E**62** (2000) 7797.
[18] H. Saleur, Nucl. Phys. B**382** (1992) 486.
[19] V. Gurarie, Nucl. Phys. B**410** (1993) 535.
[20] H. Kausch, *Curiosities at $c = -2$*, hep-th/9510149.
[21] M.R. Gaberdiel and H. Kausch, Phys. Lett. B**386** (1996) 131.
[22] F. Rohsiepe, *On reducible but indecomposable representations of the Virasoro algebra*, hep-th/9611160.
[23] V. Gurarie, M. Flohr and C. Nayak, Nucl. Phys. B**498** (1997) 513.
[24] M.R. Gaberdiel and H. Kausch, Nucl. Phys. B**538** (1999) 631.
[25] H. Kausch, Nucl. Phys. B**583** (2000) 513.
[26] E.V. Ivashkevich, J. Phys. A**32** (1999) 1691.

[27] F. Spitzer, *Principles of random walk*, 2nd edition, GTM 34, Springer, New York 1976.
[28] J. Spanier and K.B. Oldham, *An atlas of functions*, Hemisphere 1987.
[29] B.M. McCoy and T.T. Wu, *The two–dimensional Ising model*, p. 254, Harvard University Press, Cambridge 1973.